\documentclass[letter]{aastex631}
\usepackage[utf8]{inputenc}

\usepackage{multirow}
\usepackage{hhline}

\usepackage{enumerate}
\usepackage{textcomp}
\usepackage{amsmath,amssymb}
\usepackage{graphicx}
\usepackage{color}
\definecolor{orange}{rgb}{1,0.5,0}
\definecolor{purple}{rgb}{0.5,0,1}
\definecolor{darkgreen}{rgb}{0,0.5,0}
\definecolor{brown}{rgb}{0.5,0,0}
\usepackage{amsmath}

\newcommand{\beq}{\begin{equation}}
\newcommand{\eeq}{\end{equation}}
\newcommand{\ba}{\begin{array}}
\newcommand{\ea}{\end{array}}

\begin{document}

\title{Physical conditions that lead to the detection of the pair annihilation line in the BOAT GRB221009A}

\author[0000-0001-8667-0889]{Asaf Pe'er}
\affiliation{Bar-Ilan University, Ramat-Gan 5290002, Israel}

\author[0000-0002-9725-2524]{Bing Zhang}
\affiliation{The Nevada Center for Astrophysics, University of Nevada, Las Vegas, NV 89154, USA}
\affiliation{Department of Physics and Astronomy, University of Nevada, Las Vegas, NV 89154, USA}

\begin{abstract}

The brightest of all time (BOAT) GRB221009A show evidence for a narrow, evolving MeV emission line. Here, we show that this line can be explained as due to pair annihilation in the prompt emission region, and that its temporal evolution is naturally explained as the high-latitude emission (emission from higher angles from the line of sight) after prompt emission is over. We consider both the high and low optical depth for pair production regimes, and find acceptable solutions, 
with the GRB Lorentz factor $\Gamma \approx 600$ and the emission radius $r \gtrsim 10^{16.5}$~cm. 
We discuss the conditions for the appearance of such a line, and show that a unique combination of high luminosity and Lorentz factor that is in a fairly narrow range are required for the line detection. This explains why such an annihilation line is rarely observed in GRBs.

\end{abstract}

\keywords{
Relativistic jets ---  High energy astrophysics ---  Gamma-ray bursts --- Theoretical models}

\section{Introduction}

Gamma-ray bursts (GRBs) are the most extreme explosions in the universe, releasing, in many cases more than $ 10^{52} - 10^{53}$~ ergs within a few seconds. This energy release occurs in a very compact region, comparable to the core of a collapsing, massive star, namely $10^7 - 10^8$~cm \citep[see][for a few recent reviews]{Meszaros06, Peer15, KZ15, Zhang18, Meszaros19, Bosnjak+22}. Observations of very energetic photons, in the MeV- GeV (and recently, TeV) range, thus imply that the explosion energy must first be converted to kinetic energy in the form of relativistic jet before dissipation causes the release of the gamma-ray signal. Jet Lorentz factor of $\Gamma \geq 100$ is typically needed to solve the compactness problem \citep{KP91, WL95, LS01}; but see \citet{Dereli+22}. Explosion energy conversion to kinetic energy can be mediated either by photons \citep[the well known "fireball" model;][]{CR78, SP90, RM92, MR93, RM94} or by strong magnetic fields \citep{Drenkhahn02, DS02, MR97b, ZP09, ZY11}. In either scenario, a large number of $e^\pm$-pairs is expected to be created, by interaction of energetic photons having (comoving) energies exceeding the threshold energy of 0.511~MeV.

Although the creation of a large number of pairs was predicted a long time ago \citep{PW04, PMR06, Ioka+07, Murase08}, a firm observational confirmation of the existence of an expected pair annihilation line in the observed GRB spectra was so far lacking. This fact can partially be explained as that the blue-shifted line is expected to be detected in the hundred's of MeVs (the Fermi-LAT) range, in which the detector's sensitivity is poorer than at lower energy bands. In addition, due to relativistic motion, this line may be smeared if the Lorentz factor in different jet regions is somewhat different. Nonetheless, the lack of detection of this line seem to contradict with theoretical expectations.  

This situation has changed recently with the observation of GRB221009A, which became known as the 'brightest-of-all-times' (BOAT) GRB. This extremely bright burst had a reported fluence of $F \sim 0.2~{\rm erg/cm^{2}}$ \citep{An+23, Frederiks+23, Lesage+23, Burns+23}. At redshift $z=0.151$ this implies a record-breaking isotropic energy of $E_{iso} \sim 1.5 \times 10^{55}$~ erg as detected by GECAM-C \citep{An+23}, as well as a luminosity exceeding $10^{54}$~erg/s, which is also among the highest known.

This GRB was so bright that the Fermi-GBM detector was saturated in between 219-277~s from the burst trigger. 
In addition to being so bright, this burst had several other unique features. Starting at $\sim 230$~s from the trigger, extensive TeV emission was detected, with more than 5000 photons detected above TeV \citep{LHAASO23a} and the most energetic photon having an energy as high as 13~TeV \citep{LHAASO23b}. The TeV emission first rose until $\sim 242$~s, and then the flux decayed with time roughly as $t^{-1}$. Furthermore, this burst showed a clear narrow emission line at energy of $\sim 10$~MeV, separated, in both flux, energy as well as temporal behaviour from the peak of the flux which was at the 100's of keV regime. This line was first identified immediately after the saturation end time (at $\sim 280$~s), and detected for tens of seconds \citep{Ravasio+23, Zhang_Y+24}. Its energy was observed to decay in time, roughly as $\nu_{line} (t) \propto (t-226~{\rm s})^{-1}$. Both \cite{Ravasio+23} and recently \citet{Zhang_Z+24} have interpreted the line feature as due to high-latitude pair annihilation line emission. However, they did not study the detailed physical conditions to justify that a pair annihilation line with such a luminosity can be realized in this burst. 

Here, we explore the consequences of this observation. We focus on understanding the physical conditions that enable the creation of this line, as well as its uniqueness, namely why it has not been detected until recently. As we show below, a relatively narrow range of parameters is required for this line to be detected. We convincingly show that the observed signal decay is indeed due to high latitude emission, and we explore the consequences of this on the jet structure.

\section{Key observations and basic model assumptions}

The observed hard X-ray lightcurve of GRB22109A is very long, lasting for several hundreds of seconds. It is variable, with variability time of a few seconds, and show multiple peaks. The highest peak occurs at $\sim 244$~s after the GRB trigger. In fact, this peak was so bright that the Fermi-GBM detector was saturated in between the time interval 219-277~s \citep{Ravasio+23}. The physical origin of this peak is uncertain. In particular, the radius in which energy is dissipated is unknown. However, the fact that the observed spectra is broad, and extends to high energies suggests that the energy dissipation that produces this peak occurs above the photosphere. A likely possibility is that a narrow Poynting flux dominated jet dissipate energy at an emission radius beyond $10^{16}$ cm (see \cite{Dai23,ZWZ24} for arguments)\footnote{The broad-band afterglow emission requires the existence of two jet components, with a narrow core and a structured wing (see \cite{ZWZ24} for the arguments supporting such a jet structured and detailed modeling by \cite{Sato23,Zheng24,Ren24,ZhangBT24}). The prompt emission is likely related to energy dissipation within the narrow jet.} 



\begin{figure}[h]
 \includegraphics[scale=.4] {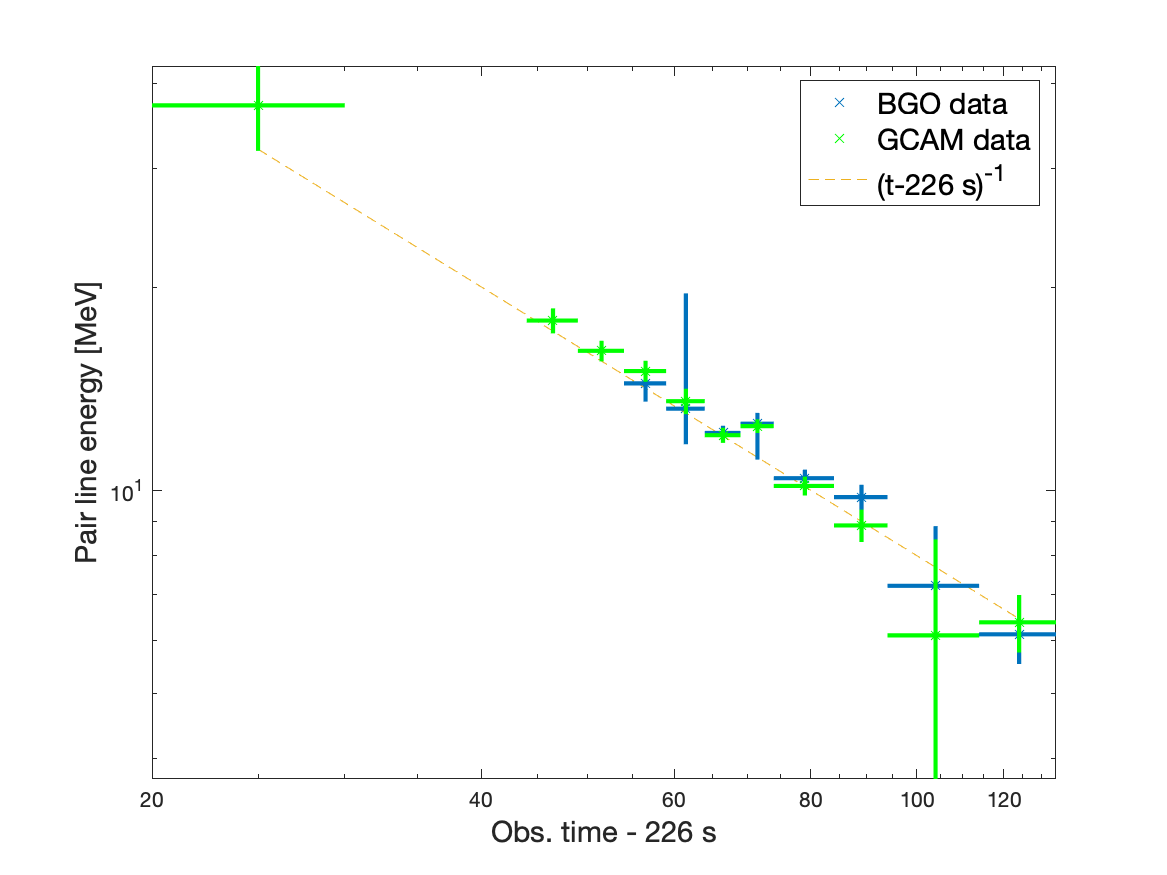}
 \includegraphics[scale=.4] {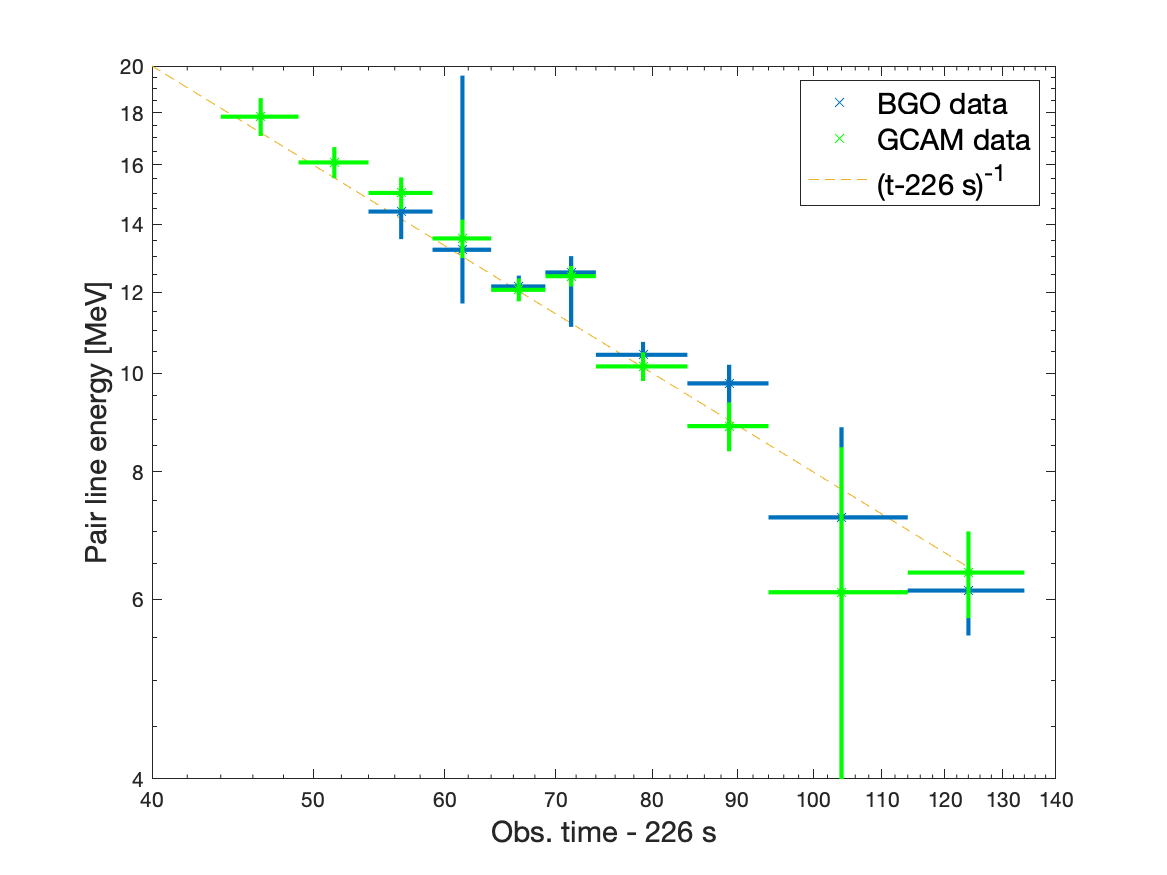}
\caption{Temporal evolution of the line energy. Left: the full data set; Right: the zoom-in version. Figure made from data taken from the tables given in \cite{Ravasio+23} and  \cite{Zhang_Y+24}.}
\label{fig:1}
\end{figure}

\begin{figure}[h]
 \includegraphics[scale=.4] {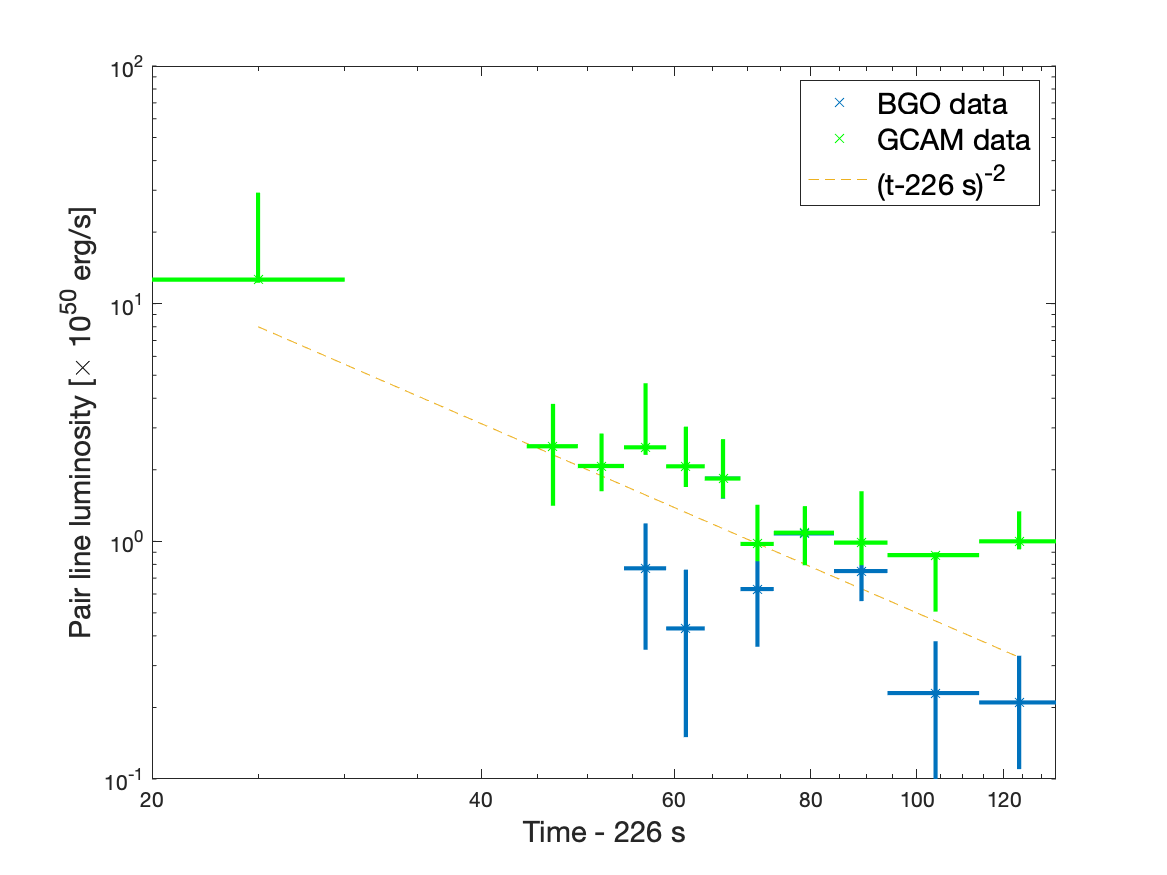}
 \includegraphics[scale=.4] {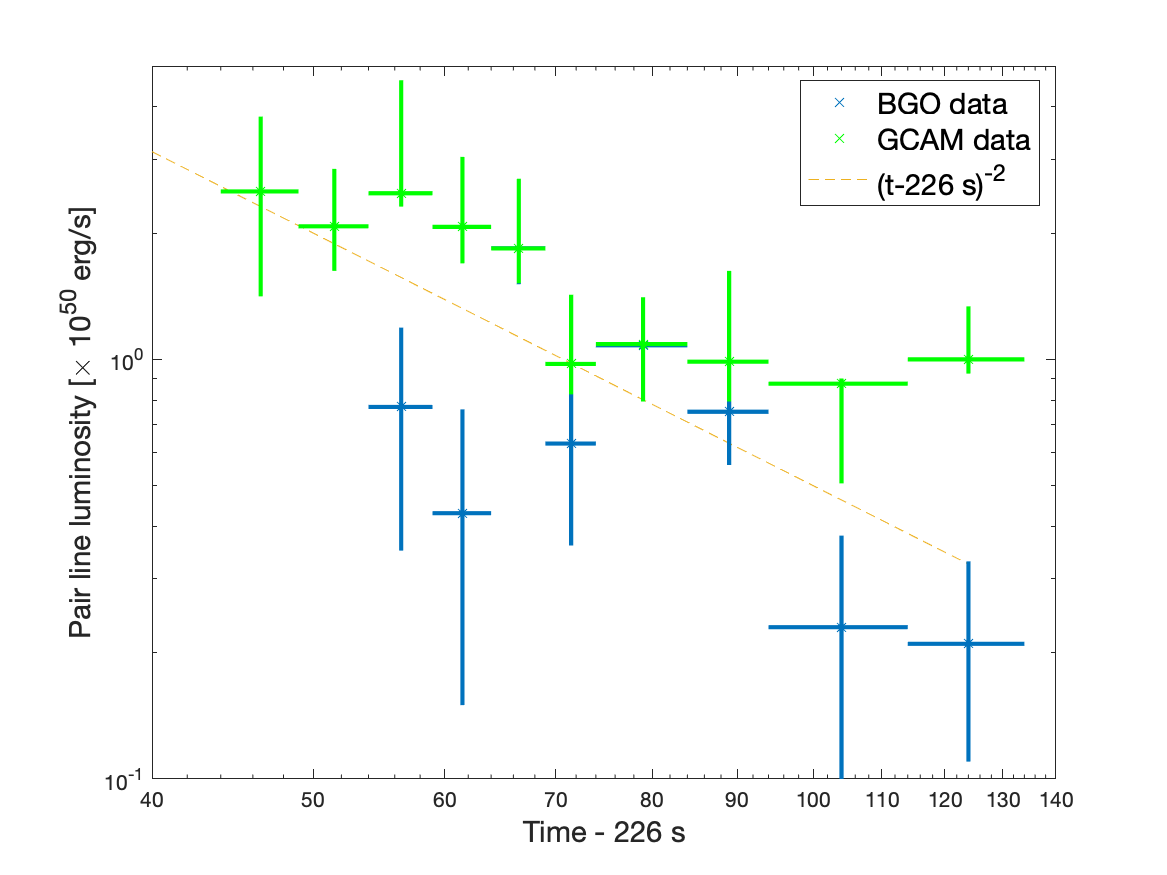}
\caption{Temporal evolution of the line luminosity. Left: the full data set; Right: the zoom-in version. Figure made from data taken from the tables given \cite{Ravasio+23} and  \cite{Zhang_Y+24}.}
\label{fig:2}
\end{figure}

After the saturation of the GBM detector, clear evidence for emission line emerges. The line energy is in the range of $\sim 10$~MeV, and decays with time very close to $h \nu_{line} (t) \propto (t-226~s)^{-1}$ (see Figure \ref{fig:1}).   The decay of the flux varies considerably, but may be  consistent with a $t^{-2}$ law (see Figure \ref{fig:2}). These scaling laws are well consistent with the prediction of high-latitude emission \citep[e.g.][]{Kumar00,ZhangBB09,Zhang18}.


As a figure of merit, the line luminosity is $L_{line}^{ob.} \approx 10^{50}$~erg/s. It is related to the comoving frame luminosity by $L_{line}^{ob.} = D^2 L'$, where $D$ is the Doppler boost factor (see below). Here and below, we use primes to describe quantities measured in the comoving frame. 
An underlying assumption of this work is that this observed line is due to pair annihilation. This assumption implies that the rate of pair annihilation must be 
\beq
\dot N'^\pm = {L' \over 2 m_e c^2} = {6 \times 10^{55} \over D^2} ~{\rm s^{-1}}.
\label{eq:1}
\eeq

For simplicity, we assume that the energy dissipation that produces the flux peak at $226$~s occurs at a single radius, $r$, which is the radius when the prompt emission abruptly stops. To estimate this radius, we use the characteristic observed lightcurve variability time scale, $\Delta t \approx 10$~s. This gives 
\beq
r \simeq \Gamma^2 c \Delta t^{ob.} = 1.08 \times 10^{17}~\left({\Gamma \over 600}\right)^2 \Delta t_1~{\rm cm}. 
\label{eq:2}
\eeq
Here and below, we estimate a fiducial value to the Lorentz factor of  $\Gamma = 600$ \citep{ZWZ24} for the calculation. In section \ref{sec:high_lat} below, we compare the results with the data, to constrain the value of the Lorentz factor. We further use the standard notation $Q_x \equiv Q/10^x$ in cgs units. Note that this calculation takes into account both the radial and side spreading of the expanding plasma shell.

As the observed spectrum is flat, the flux of energetic photons in the plasma is high. As a result, a large number of pairs are produced. The produced pairs rapidly cool, mainly radiatively, as well as adiabatically. The cold pairs annihilate, producing the observed line. The observed line energy drops with time due to the decrease of the Doppler factor of high latitude emission.

Using the observed (total) burst luminosity  $\sim 10^{54}~{\rm erg~s^{-1}}$, one can estimate the photospheric radius to be at 
\beq
r_{ph} = {L \sigma_T \over 8 \pi m_p c^3 \Gamma^3} = {5.8 \times 10^{20} \over \Gamma^3}~L_{54} ~{\rm cm}.
\label{eq:3}
\eeq
Thus, for jet Lorentz factor of a few hundreds, the assumed radius of pair emission line is much greater than the photosphere. However, this results strongly depends on the value of the jet Lorentz factor, $\Gamma$.

\section{Theoretical expectation of the pair annihilation rate}

The line luminosity is due to annihilation of pairs. In order to estimate the expected luminosity, one needs to calculate the rate of pair annihilation. Here we calculate the theoretical expectation of this rate under various plausible plasma conditions.

\subsection{Pair production rate}

The pair annihilation rate depends on the density of pairs inside the plasma. As pairs are both produced and annihilated, their density is set by the balance between these two processes.  Thus, we first   
calculate the rate of pair production inside the plasma, from which one can deduce the pair density.

Calculation of the pair production rate is done by estimating the density of photons inside the plasma that are energetic enough to produce pairs. 
The comoving number density of photons having energies exceeding 
$\epsilon' \geq A m_e c^2$ is calculated assuming an approximately flat spectra ($\nu F_\nu \propto \nu^0$), which gives an average photon energy $\langle \epsilon'
\rangle \approx A m_e c^2$. Here, $A\geq 1$ is an uncertainty factor, which takes into account deviation from a flat spectrum, variation of the cross section with energy, etc. Its value could be much greater than unity but could be of order of unity, depending on the physical parameters (see more discussion below).
For a flat spectrum (which is suggested by both Fermi and GECAM spectral data), the number density of photons in the plasma is therefore  
\beq
n'_\gamma(\epsilon' = A m_e c^2) = { L \over 4 \pi A m_e c^2 \Gamma^2 c r^2} = {7.7 \times 10^8 \over A}~ L_{54} ~ \Delta t_{1}^{-2} ~\left({\Gamma \over 600}\right)^{-6}~{\rm cm^{-3}}.
\label{eq:4}
\eeq 

The energetic photons interact with low energy photons to produce electron-positron pairs. For a photon with (comoving) energy $A m_e c^2$, the cross section for pair production is maximal for producing a pair with a photon of (comoving) energy $\epsilon' \approx A^{-1} m_e c^2$, and is equal to $\approx \sigma_T$.  
The typical (comoving) timescale for producing pairs is therefore 
\beq
(t')^{-1} = n'_\gamma (\epsilon')  c \sigma_T = { L \sigma_T \over 4 \pi A^{-1} m_e c^2 \Gamma^2 r^2} = 1.5 \times 10^{-5} A ~ L_{54} ~ \Delta t_{1}^{-2} ~\left({\Gamma \over 600}\right)^{-6}~{\rm s^{-1}},
\label{eq:5}
\eeq
or $t' = 6.5 \times 10^4$~s. This value should be compared to the comoving (dynamical) time, $t'_{dyn} = r/ \Gamma c = \Gamma \Delta t =  6 \times 10^3~(\Gamma/600) \Delta t_1$~s. The fact that $t' \gg t'_{dyn}$ (for the fiducial parameter values chosen) implies that for these parameter values, only a small fraction of the photons will produce pairs within the available (dynamical) time.

The comoving width of the expanding plasma is $\Delta r' = r/\Gamma$. 
The optical depth for producing pairs is thus 
\beq
\tau_{\gamma \gamma} = \Delta r' n'_\gamma(\epsilon') \sigma_T = {L \sigma_T \over 4 \pi A^{-1} m_e c^3 \Gamma^3 r} = 0.092 A ~ L_{54} ~ \Delta t_{1}^{-1} ~\left({\Gamma \over 600}\right)^{-5}
\label{eq:6}.
\eeq

Given the high uncertainty in the value of the jet Lorentz factor and the very strong dependence of the optical depth on its unknown value, there is a large uncertainty in the value of the optical depth, $\tau_{\gamma \gamma}$. Since a qualitatively different result is expected for high ($\tau_{\gamma \gamma} > 1$) and low ($\tau_{\gamma \gamma} < 1$) optical depths, 
we split the calculations into these two regimes: low optical depth and high optical depth for pair production.

\subsection{Low optical depth for pair production }

For optical depth $\tau_{\gamma \gamma} < 1$, the intensity drops as $I(\tau_{\gamma \gamma}) \simeq I_0 (1 - \tau_{\gamma \gamma})$. This means that only a fraction $\tau_{\gamma \gamma}$ of the photons interact to produce pairs before escaping the plasma. In this regime, no sharp cutoff in the high energy spectra that results from high energy photon annihilation is expected. Rather, some gradual decay in the high energy spectra is predicted. We therefore expect in this regime that the value of the uncertainty parameter $A$ will not be too far from unity.

The rate of pair production by a single photon (number of pairs produced per second) is given by $n'_\gamma (\epsilon') c \sigma_T$. Therefore, the density of the pairs produced inside the plasma per unit time is given by 
\beq
\ba{lcl}
\dot n'_{e^\pm} & = & n'_\gamma (\epsilon') n'_\gamma (\epsilon') c \sigma_T \\
& = & {L^2 \over (4 \pi A^{-1} m_e c^2 \Gamma^2 c r^2)^2 } c \sigma_T = 1.2 \times 10^4 A^2~  L_{54}^2 ~ \Delta t_{1}^{-4} ~\left({\Gamma \over 600}\right)^{-12} ~{\rm cm^{-3}~ s^{-1}},
\ea
\label{eq:7}
\eeq
which is given in units of number per unit volume per unit time. Note the extreme dependence on the values of the parameters, especially the Lorentz factor, $\Gamma$. 

Assuming that this rate is constant during the dynamical time, $t'_{dyn} = r/\Gamma c$, the density of pairs inside the plasma at the end of the dynamical time is given by 
\beq
n'_{e^\pm} = {A^2 L^2 \sigma_T \over (4 \pi)^2 (m_e c^2)^2 \Gamma^5 c^2 r^3}.
\label{eq:8}
\eeq

\subsubsection{Density of pairs inside the plasma: low optical depth}
We assume that the pairs that are produced rapidly cool, reaching typical velocity $v < c$. In this case, the cross section for pair annihilation is $v \sigma_{ann.} \simeq c \sigma_T$. The optical depth for pair annihilation is therefore 
\beq
\tau_{e^\pm} = \Delta r \sigma_T n'_{e^\pm} = {L^2 \sigma_T^2 \over (4 \pi)^2 (m_e c^2)^2 \Gamma^6 c^2 r^2} = 8.5 \times 10^{-3} A^2~  L_{54}^2 ~ \Delta t_{1}^{-2} ~\left({\Gamma \over 600}\right)^{-10},
\label{eq:9}
\eeq 
where $\Delta r = r/\Gamma$ was used. 
This optical depth must be smaller than unity, since $n'_{e^\pm} < n'_\gamma$ (this is indeed seen in the equation). Therefore, the rate of annihilation is 
\beq
\dot n'_{ann} = (n'_{e^\pm})^2 c \sigma_T.
\label{eq:10} 
\eeq 

We may therefore write the rate equation for the time evolution of pair density inside the plasma:
\beq
\dot n'_{e^\pm} = (n'_\gamma)^2 c \sigma_T -  (n'_{e^\pm})^2 c \sigma_T  \equiv C_1 - C_2 (n'_{e^\pm})^2
\label{eq:11}
\eeq
(where $C_1$ and $C_2$ are constants). The solution to this equation is 
\beq
n'_{e^\pm} = n'_\gamma \tanh \left[(n'_\gamma c \sigma_T) t \right].
\label{eq:12}
\eeq
Here, $n'_\gamma = \sqrt{n'_\gamma(\epsilon) n'_\gamma (\epsilon')}$. 

This means that in a steady state, 
\beq
n'_{e^\pm} \underset{t\rightarrow \infty}{\rightarrow} n'_\gamma = 7.7 \times 10^8 A~L_{54}~\Delta t_{1}^{-2} ~\left({\Gamma \over 600}\right)^{-6}~{\rm cm^{-3}},
\label{eq:13}
\eeq
and that the characteristic timescale for approaching this solution is 
\beq
t'_{ss} =  {1 \over n'_\gamma c \sigma_T } = {4 \pi A^{-1} m_e c^2 \Gamma^2 r^2 \over L \sigma_T} =  6.5 \times 10^4 A^{-1}~  L_{54}^{-1} ~ \Delta t_{1}^{2} ~\left({\Gamma \over 600}\right)^{6}~{\rm s}.
\label{eq:14}
\eeq
(Note again that all times are measured in the comoving frame). 
This time should be compared to the  (comoving) dynamical time, $t'_{dyn} = r/\Gamma c =  6 \times 10^3~(\Gamma/600) \Delta t_1$~s. One thus concludes: (i) that the time to reach steady state is about an order of magnitude longer than the available time; and (ii) this time depends on $\Gamma$. The condition for $t'_{ss} > t'_{dyn}$ is translated into 
\beq
\Gamma > 370 A^{1/5}~  L_{54}^{1/5} ~ \Delta t_{1}^{-1/5}. 
\label{eq:15}
\eeq
For lower values of $\Gamma$, the density of pairs inside the plasma approaches the asymptotic value given in Equation \ref{eq:13} (with an inverse dependence on $\Gamma$!). For a larger Lorentz factor, a steady state is not reached. 

We proceed by assuming that $t'_{ss} \gg t'_{dyn}$. In this case, one can approximate $\tanh(x) \approx x$, which is valid for $x \ll 1$. Equation \ref{eq:12} then gives the comoving pair density inside the plasma, 
\beq
\ba{lcl}
n'_{e^\pm}(t')  & \approx  & n'_\gamma (n'_\gamma c \sigma_T) t'  =  (n'_\gamma)^2 c \sigma_T t' \\
& = & {L^2 \sigma_T \over (4 \pi)^2 A^{-1} (m_e c^2)^2 \Gamma^4 c r^4} t'= 1.18 \times 10^4 t' ~  A L_{54}^2 ~ \Delta t_{1}^{-4} ~\left({\Gamma \over 600}\right)^{-12} ~{\rm cm^{-3}}.
\ea
\label{eq:16}
\eeq
Note the dependence on the uncertain factor of $A$ is linear.

We may now calculate the rate of annihilation, namely the total number of photons in the emission line that are emitted per unit comoving time (second) from the entire comoving volume,
\beq
\dot N'_{ann} = \dot n'_{ann} V' = (n'_{e^\pm})^2 c \sigma_T V'
\label{eq:17}
\eeq
(see Equation \ref{eq:10}). Here, $V' = 4 \pi r^2 c t_{dyn} = 2.6 \times 10^{49} ~ (\Gamma/600)^5 \Delta t_1^3~{\rm cm^3}$ is the comoving volume\footnote{We note that we assume here a constant comoving volume. The error introduced by this assumption is of the order unity.}.
Putting numbers, one finds:
\beq
\dot N'_{ann} = {L^4 \sigma_T^3 \over (4 \pi)^3 A^{-2} (m_e c^2)^4 \Gamma^{9} c r^{5}} t'^2 = 7.4 \times 10^{43} t'^2 ~ A^2 L_{54}^4 ~ \Delta t_{1}^{-5} ~\left({\Gamma \over 600}\right)^{-19} ~{\rm s^{-1}}.
\label{eq:18}
\eeq
At $t'=t'_{dyn}$, this reads 
\beq
\dot N'_{ann}(t'_{dyn}) = { L^4 \sigma_T^3 \over (4 \pi)^3 A^{-2} (m_e c^2)^4 \Gamma^{11} c^3 r^{3}}  = 2.7 \times 10^{51} A^2 ~ L_{54}^4 ~ \Delta t_{1}^{-3} ~\left({\Gamma \over 600}\right)^{-17} ~{\rm s^{-1}}.
\label{eq:19}
\eeq
One therefore finds a huge dependence of the pair annihilation rate on the value of the Lorentz factor, $\Gamma$. This dependence enables a very wide range of values to be achieved. However, this result holds only as long as the condition in Equation \ref{eq:15} is met, otherwise the optical depth becomes larger than unity, and the parametric dependence becomes different.

It is easy to show that the condition $t'_{ss} > t'_{dyn}$ is equivalent to the condition $\tau_{\gamma \gamma} < 1$.  Thus, if the conditions are such that $t'_{ss} < t'_{dyn}$, then $\tau_{\gamma \gamma} > 1$, and one needs to consider the regime of high optical depth, discussed below. 



\subsection{High optical depth for pair production}

We here explore the other regime, in which the optical depth for pair production, $\tau_{\gamma \gamma} > 1$ (see Equation \ref{eq:6}). The condition $\tau_{\gamma \gamma} > 1$ is equivalent to the condition that the comoving time for producing pairs (Equation \ref{eq:5}) is smaller than the dynamical time, i.e., 
$$
\tau_{\gamma \gamma} > 1 ~ \leftrightarrow t' < t'_{dyn}. 
$$
Therefore, in this regime, production of pairs occurs on a timescale shorter than the dynamical time.  The pair production rate can therefore be estimated as the rate in which energy is produced in photons having energies  $\epsilon'_{ph} \geq A m_e c^2$.
This rate (per unit volume) is simply 
\beq
\dot n'_{e^\pm} \approx {L \over (4 \pi r^2 \Gamma^2 c)} {1 \over A m_e c^2 t'_{dyn}}.
\label{eq:20} 
\eeq 

The produced pairs rapidly cool. 
At low energies, the cross section for pair annihilation is also $\sim \sigma_T$. Therefore, the rate of pair annihilation is $\dot n'_{ann} \sim {n'_{e^\pm}}^2 c \sigma_T$. For large  optical depth $\tau_{\gamma \gamma} \gg 1$, there is a balance between pair production and pair annihilation rates, namely  $\dot n'_{e^\pm} = \dot n'_{ann}$, giving
\beq
{n'_{e^\pm}}^2 =  {L \over (4 \pi r^2 \Gamma^2 A m_e c^3 t'_{dyn})} {1 \over c \sigma_T} = {L \sigma_T t'_{dyn}  \over (4 \pi r^2 \Gamma^2 A m_e c^2)} \left({1 \over c \sigma_T t'_{dyn}}\right)^2 = {\tau_{\gamma \gamma}  \over A^2 (c \sigma_T t'_{dyn})^2},
\label{eq:21}
\eeq
where we made use of $r \sim \Gamma c t'_{dyn}$. 

One therefore finds that in this regime of high optical depth to pair production, the number density of pairs inside the plasma is expected to be 
\beq
n'_{e^\pm} \sim {\tau_{\gamma \gamma}^{1/2} \over A c \sigma_T t'_{dyn}} = 2.5 \times 10^9~A^{-1/2} ~ L_{54}^{1/2} ~ \Delta t_{1}^{-3/2} ~\left({\Gamma \over 600}\right)^{-7/2}~{\rm cm^{-3}}.
\label{eq:22}
\eeq  
This result further implies that the optical depth to pair annihilation is 
\beq
\tau_{e^\pm} = \Delta r n'_{e^\pm} \sigma_T = \tau_{\gamma \gamma}^{1/2} A^{-1},
\label{eq:23}
\eeq
namely, at any given radius, $\tau_{\gamma \gamma}$ and $\tau_{e^\pm}$ are either both greater or smaller than unity (for $A=1$). 

The rate of annihilation, namely the total number of photons in the emission line that are emitted per unit comoving time from the entire comoving volume, can now be calculated using equation (\ref{eq:20}),
\beq
\ba{lcl}
\dot N'_{ann} = \dot n'_{ann} V' = \dot n'_{e^\pm} V' & = & {L \over (4 \pi r^2 \Gamma^2 c)} {1 \over A m_e c^2 t'_{dyn}} \times 4 \pi r^2 c t'_{dyn}  =  {L \over \Gamma^2 A m_e c^2} \\ & = & 3.4 \times 10^{54}~A^{-1}~L_{54} ~  \left({\Gamma \over 600}\right)^{-2}~{\rm s^{-1}}.
\ea
\label{eq:24}
\eeq
Note that essentially, this result means that the brightness of the line is equal to the rate of energy loss from high energy photons by annihilation.

\begin{figure}[h]
 \includegraphics[scale=.8] {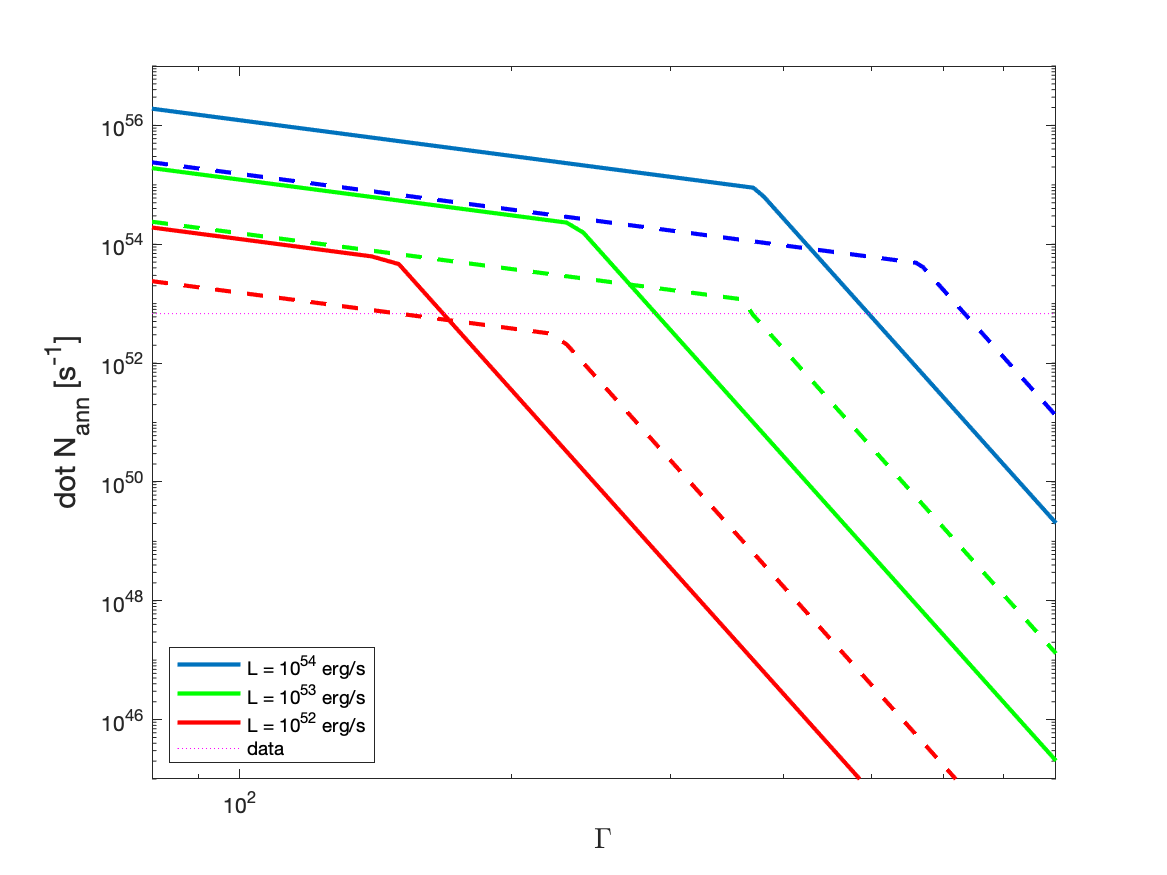}
\caption{Rate of pair annihilation as a function of the Lorentz factor, $\Gamma$, from Equations \ref{eq:19} and \ref{eq:24}. At low $\Gamma$ the system is in the high optical depth to pair production regime, and the rate of annihilation is proportional to $\Gamma^{-2}$, while at higher values of $\Gamma$ (the low optical depth regime), it sharply decays as $\Gamma^{-17}$. Solid lines are for $A=1$, while dashed lines are for $A=8$ (note that $A=10$ would result in lines overlap).} 
\label{fig:3}
\end{figure}

The rate of pair annihilation, as calculated in Equations (\ref{eq:19}) and (\ref{eq:24}) for the low and high optical depth regimes is plotted in Figure \ref{fig:3}, for different luminosities, together with the inferred value from the data. In the plot, observed variability time $\Delta t = 10$~s was assumed. Values of $A=1$ (solid lines) and $A=8$ (dashed lines) were taken for demonstration. This helps explaining the rareness of this line; see further discussion below.  

\section{The Doppler boost: high latitude emission}
\label{sec:high_lat}

The observed line show a clear temporal decay, with peak energy that is well fitted as $\epsilon_{\rm line} = h \nu_{\rm line} \propto (t-226~s)^{-1}$. This temporal behaviour is typical for emission originating at high-lattitude, and is due to the evolving Doppler boost. 

The comoving pair annihilation line energy ($\epsilon' = 0.5$~MeV) is Doppler boosted towards the observer: $\epsilon^{ob.} = D \epsilon'$. Here,
$$
D = {1 \over \Gamma(1-\beta \cos \theta)},
$$ 
where $\theta$ is the angle to the line of sight. Under the simplified hypothesis we use here, all the photons are emitted simultaneously at the same radius, $r$, but at different angles to the line of sight.  Since photons emitted at angle $\theta$ are delayed with respect to photons emitted on the line of sight by an observed time 
$$
\delta t^{ob.} = {r \over c} (1 - \cos \theta) \approx {r \over c} {\theta^2 \over 2}, 
$$
one finds the Doppler boost to be 
\beq
\ba{l}
\theta < {1 \over \Gamma} \rightarrow D \simeq 2 \Gamma \propto {\delta t^{ob.}}^0,\\
\theta > {1 \over \Gamma} \rightarrow D \simeq {2 \over \Gamma \theta^2} = {r \over \Gamma c \delta t^{ob.}}. 
\ea
\label{eq:25}
\eeq
Note that here we use the symbol $\delta t^{ob.}$ to describe the difference between the observed time of the first photon (emitted on the line of sight) and a photon emitted at angle $\theta$ to the line of sight, while $\Delta t$ is used to describe the characteristic observed variability time, from which the emission radius is estimated.  

One may now compare this theoretical expectation to the data. About $\delta t^{ob.} \simeq 60$~s after the peak of the flux, the pair annihilation line is at $\sim 15$~MeV, namely $D \simeq 30$. Using $r = \Gamma^2 c \Delta t$, one finds 
$ r/(\Gamma c) = \Gamma \Delta t = 1800$, or $(\Gamma/600) \Delta t_1 = 0.3$. 
This result can be combined with the result in Equation (\ref{eq:25}) (bottom) inside Equation (\ref{eq:1}), using observed time $\delta t^{ob.} = 60$~s, to write the observational constraint in Equation (\ref{eq:1}) as a constraint on the rate of pair annihilation at the end of the dynamical time, 
\beq
\dot N^\pm = {L' \over 2 m_e c^2} = {L^{ob.} \over 2 m_e c^2}\left({\delta t^{ob.} \over \Gamma \Delta t} \right)^2 = 6.8 \times 10^{52}~s^{-1}.
\label{eq:26}
\eeq
where we took $L^{ob.} = 10^{50}$~erg/s as the luminosity of the line at $\delta t^{ob.} = 60$~ from the peak time.
  
\subsection{Low optical depth for pair production}  
  
The pair annihilation rate calculated in Equation (\ref{eq:26}) is to be compared to the theoretical expectation given in Equation (\ref{eq:19}). Basic algebra reveals that the two results match for $\Gamma = 620~L_{54}^{2/7}$ and $\Delta t = 3.0$~s. Assuming these value of the Lorentz factor, the dissipation radius is $r = 3.5 \times 10^{16}$~cm, which is much higher than the photospheric radius, at $r_{ph} = 2.4 \times 10^{12}$~cm. The optical depth for pair production is $\tau_{\gamma \gamma} = 0.26~A$, i.e., close to unity, and for pair annihilation it is $\tau_{e^\pm} = 0.068~A^2$. For these parameters, The time to reach steady state, $t'_{ss} = 7.1 \times 10^3$~s is indeed longer than the dynamical time, $t'_{dyn} = 1.86 \times 10^3$~s (although by a factor of a few only), implying that the approximation we used above is valid.

\subsection{High optical depth for pair production}  

An alternative scenario may be that the emission occurs in region of high optical depth to pair production. In this scenario, the constraint set by the data (Equation \ref{eq:26}) is to be compared to the theoretical expectation as given in Equation (\ref{eq:24}).

The two results match, provided that  $(\Gamma/600) = 7~L_{54}^{1/2} A^{-1/2}$. However, too high a value of $\Gamma$ is unacceptable, as in this case the optical depth for pair production (Equation \ref{eq:6}) will be smaller than unity. Using the constraint $(\Gamma/600) \Delta t_1 = 0.3$ inside  Equation (\ref{eq:6}) leads to the constraint
$$
0.3 A L_{54} >  \left({\Gamma \over 600}\right)^{4},
$$
which, combined with the constraint on the line luminosity above, results in the condition 
\beq
A > 20~ L_{54}^{1/3}.
\label{eq:27}
\eeq
This is translated to an expected spectral cutoff at energy 
\beq
\epsilon^{ob.} > 20 D m_e c^2 = 20 \left({1800~s \over t^{ob.}} \right) m_e c^2 = 10 \left({1800~s \over t^{ob.}} \right)~{\rm MeV}.
\label{eq:28}
\eeq
this regime is not excluded by existing data, and is therefore valid as well. 

The cross section for photon scattering by the created pairs in the plasma, $\sigma_T$,  is comparable to the cross section of annihilation of cold pair. Therefore, Equation \ref{eq:23} provides a very good proxy for the optical depth for scattering by the created pairs. Using Equation \ref{eq:6} and the observational constraint set above, the optical depth for photon scattering is given by 
\beq
\tau_{sc.} = 0.55~A^{-1/2} ~ L_{54}^{1/2}  ~\left({\Gamma \over 600}\right)^{-2} < 0.12  ~ L_{54}^{1/3}  ~\left({\Gamma \over 600}\right)^{-2}, 
\label{eq:tau_sc}
\eeq
where in the last equality we used equation \ref{eq:27}.
One therefore finds that for most of parameter space regime of interest, this optical depth is smaller than unity, and even for the most extreme cases of $L = 10^{54}$~erg/s and $\Gamma \gtrsim 100$, it is not more than a few at most. (Lower values of Lorentz factor will imply that the emission radius is close to the photospheric radius, and the resulting spectra will become quasi-thermal. The observed spectra therefore exclude this parameter space). One therefore concludes that scattering by the pairs is not expected to have a major effect on the observed line properties, such as its observed brightness or width.

\section{Conclusions and Discussion}

\subsection{Pair annihilation as a source of the observed line}

In this paper, we considered the narrow emission line seen in GRB221009A. We argue that this line originates from $e^\pm$ pair annihilation. We calculated the expected rate of pair annihilation, as a function of the observed GRB parameters, namely the luminosity, a spectral cutoff ($A$) and variability time, and the unknown jet Lorentz factor. We calculated this rate in the two regimes: low and high optical depths for pair production. The results are presented in Equations (\ref{eq:19}) and (\ref{eq:24}), respectively. 

When comparing the theoretical expectations to the observed line luminosity, we found that both scenarios are plausible, and both lead to high Lorentz factor, of $\sim 600$. The reason why such two different scenarios can provide acceptable explanations is the uncertainty in the cutoff energy, $A m_e c^2$. Therefore, observational discrimination between the two scenarios is possible, by detecting a spectral cutoff ($A$). For the available data of GRB221009A, this cutoff is expected in the sub-GeV range (Equation \ref{eq:27}). This calculation can be trivially generalized to the data of any other source.

\subsection{High latitude emission}

We explain the observed line energy, as well as the temporal decay of this line as due to emission from high-latitude after the line-of-sight prompt emission is over, i.e., at high angles to the line of sight. This interpretation naturally explains why this line is detected only in the 10~MeV range, rather than the naive expectation of $\Gamma m_e c^2 \sim 300$~MeV range,
as would be expected had the emission originated from the line of sight. Indeed, according to our model, at very early times, comparable to $\Delta t \approx 3$~s after the peak (at 226~s), the Doppler boost is expected to be $2 \Gamma = 1200$, and the emission line is expected at the sub-GeV range. However, during this time, the detectors are saturated. 

Furthermore, this interpretation provides a natural explanation to the temporal decay of the line frequency as $(t-226~s)^{-1}$, and its luminosity decay as $(t-226~s)^{-2}$, although this is less pronounced by the data. 
This high latitude interpretation further indicates that the dissipation episode that leads to the emission of this line is instantaneous -- it occurs over a short observed period of a few seconds, rather than being continuous. Furthermore, the data clearly indicates that the energy dissipation episode occurs far above the photosphere, and therefore is not directly related to the on-going events that take place in the central engine during the prompt phase.   

Indeed, the high latitude emission model has a firm prediction. Assuming a $\delta$-function emission in time and radius (but that extends over a wide angular range), the observed signal should be constant (time-independent) for a duration of $\Delta t^{ob.} \approx R/(2\Gamma^2 c)$, and after that the flux should drop as $\Delta t^{-2}$. This is a universal prediction, which is independent on the origin of the emission, provided it occurs at a constant radius.

This prediction aligns very well with the observed properties of the narrow emission line in GRB221009A. However, this only gives an upper limit on the decay rate of a given signal. Indeed, other parts of the spectra do not follow this behaviour. For example, the peak of the flux seen by Fermi and GECAM is much more variable, and show high peaks at later times, up to 600~s. This indicates that multiple energy dissipation episodes occurred, though none of them provides sufficient energy to produce an observed pair signal. 

The TeV flux shows a very different temporal behaviour. While it starts its rise at the same time, 226~s from the trigger, it rises for $\approx 20$~s, before decaying as $F_{\rm TeV} \propto t^{-1.1}$, which is much shallower than the limit set by the high latitude emission. The shape of lightcurve is consistent with synchrotron self-Compton scattering from the external shock \citep{LHAASO23a}. Since the TeV photons are observed, they are not attenuated to produce pairs, likely because the emission radius is larger than the prompt emission radius where $\tau_{\gamma\gamma}$ is much smaller.




\subsection{Rarity of this line}
The analysis carried here can be used to explain the lack of detection of pair annihilation line in GRBs so far, despite the clear theoretical prediction. 
One characteristic of this particular GRB (221009A) is its extreme brightness, $L \simeq 10^{54}$~erg/s \citep{Burns+23}. This is about two - three orders of magnitude brighter than the average GRB.  

Assume that for less bright GRBs the detectors would not saturate, continuous observations would occur during the dynamical time. Thus, one could approximate the Doppler boost as $D \simeq 2 \Gamma$. The observed line luminosity is expected to be 
$$
L_{line}^{ob.} = D^2 L' = 4 \Gamma^2 \times 2 m_e c^2 \dot N^\pm.
$$ 
The results of Equations (\ref{eq:19}) and (\ref{eq:24}) provide the theoretical expectation, 
\beq
L_{line}^{ob.} = \left\{ \ba{ll} 
6.2 \times 10^{51}~~ L_{54}^4 ~ \Delta t_{1}^{-3} ~\left({\Gamma \over 600}\right)^{-15} ~~{\rm erg/s} & ({\rm low~ optical~ depth}) \\
7.8 \times 10^{54}  ~A^{-1}~L_{54}~~ {\rm erg/s} & ({\rm high~ optical~ depth}). 
\ea \right.
\label{eq:29}
\eeq

For typical GRB with luminosity $L\sim 10^{51}$~erg/s, the transition between high and low optical depth occurs at  $\Gamma = 90~L_{51}^{1/5}~\Delta t_1^{-1/5}$ (see Equation \ref{eq:15}). Thus, for typical values of $\Gamma \geq 100$, the valid regime is  always the low optical depth regime, where the line luminosity depends on the GRB luminosity to the power 4. In this regime, the theoretical calculations show a huge dependence on the value of $\Gamma$, $L_{line}^{ob.} \propto \Gamma^{-15}$. This dependence implies that $\Gamma$  must have a narrow range of values in order for the line to be detected. E.g., for $L = 10^{51}$, $\Gamma > 150$ will produce a too dim emission line.

For brighter GRBs, this constraint is somewhat releases. For example, for $L = 10^{52}$~erg/s, annihilation line from GRBs with Lorentz factor of 200-250 should be within the detected range, at least for a typical time comparable to the variability time, $\Delta t$. In such cases, this line is expected in the same energy range as the line seen in GRB221009A. However, for such a moderate luminosity GRB located at a typical cosmological distance, the line would have too few photons to be detected.

\subsection{The dissipation radius}

One of the most interesting results of our analysis is that the radius in which energy is dissipated to produce the observed line is large, $r_{dis} \sim 10^{16.5}$~cm.  This radius is far above the photospheric radius. On the other hand, this radius cannot be associated with the onset of the afterglow phase, as the data indicates that the prompt emission phase lasts much longer. 

The origin of the dissipation process at this radius is therefore not entirely clear. Due to the very large amount of energy released in this radius, We find internal shocks to be a less appealing scenario, as only the differential energy can be released, and as a result the efficiency of energy conversion is very low, typically a few \%. The two alternative options are therefore either dissipation of a large amount of magnetic energy \citep[e.g.,][]{ZY11,ZWZ24}, or alternatively, an abrupt change in the external conditions. As was shown recently \citep{PR24}, this can naturally occur by interaction of the jet with the contact discontinuity that is naturally expected close to the edge of the wind bubble produced by the progenitor stellar wind.

\section*{Acknowledgements} 
AP acknowledges the support from the European Union (EU) via ERC consolidator grant 773062 (O.M.J.). BZ acknowledges NASA 80NSSC23M0104 and the Nevada Center for Astrophysics for support.

\bibliography{references}{}
\bibliographystyle{aasjournal}

\end{document}